\documentclass[final,3p,times,onecolumn]{elsarticle}
\usepackage{graphicx}
\usepackage{longtable} 
\usepackage{lineno,hyperref}
\usepackage{booktabs} 
\usepackage{graphicx}
\usepackage{multirow}
\usepackage{natbib}
\usepackage[table,xcdraw]{xcolor}

\begin{document}
\begin{frontmatter}

\title{Twitter Permeability to financial events: an experiment towards a model for sensing irregularities}

\author[1]{Ana Fern\'andez-Vilas\corref{c1}}
\ead{avilas@det.uvigo.es}
\author[1]{Rebeca P. D\'{\i}az Redondo}
\ead{rebeca@det.uvigo.es}

\author[2]{Keeley Crockett}
\ead{k.crockett@mmu.ac.uk}

\author[2]{	Majdi Owda}
\ead{m.owda@mmu.ac.uk}

\author[2]{Lewis Evans}
\ead{L.Evans@mmu.ac.uk}

\affiliation[1]{organization={atlanTTic, I\&C Lab},
   addressline={Escuela de Ingeniería de Telecomunicación. Campus universitario s/n}, 
    city={Vigo},
    postcode={36310},
    country={Spain}
}
\affiliation[2]{organization={School of Computing, Mathematics \& Digital Technology. Manchester Metropolitan University},
    addressline={}, 
    city={Manchester},
    postcode={M1 5GD},
    country={UK}}

\begin{abstract}
	There is a general consensus of the good sensing and novelty characteristics of Twitter as an information media for the complex financial market. This paper investigates the permeability of Twittersphere, the total universe of Twitter users and their habits,	towards relevant events in the financial market. Analysis shows that a 	general purpose social media is permeable to financial-specific events	and establishes Twitter as a relevant feeder for taking decisions regarding the financial market and event fraudulent activities in that market. However, the provenance of contributions, their different levels 	of credibility and quality and even the purpose or intention behind them should to be considered and carefully contemplated if Twitter is used as a single source for decision taking. With the overall aim of this research, to deploy an architecture for real-time monitoring of irregularities in the financial market, this paper conducts a series of 	experiments on the level of permeability and the permeable features of 	Twitter in the event of one of these irregularities. To be precise, Twitter data is collected concerning an event comprising of a specific 	financial action on the 27th January 2017:{~ }the announcement about the merge of two companies Tesco PLC and Booker Group PLC, listed in the main market of the London Stock Exchange (LSE), to create the UK's	Leading Food Business. The experiment attempts to answer five key 	research questions which aim to characterize the features of Twitter permeability to the financial market. The experimental results confirm 	that a far-impacting financial event, such as the merger considered,	caused apparent disturbances in all the features considered, that is, information volume, content and sentiment as well as geographical provenance. Analysis shows that despite, Twitter not being a specific financial forum, it is permeable to financial events.

\end{abstract}


\end{frontmatter}

\section{Introduction}
\label{sec:1}

Progressive usages of technology in the stock markets have led to a
continuous growth in their business. Businesses and individual investors  alike can take decisions and invest within minutes if not in seconds \cite{Hobijn2001}. Helping both businesses and individual  investors harvest information from diverse sources such as the company  itself, through their website or Regulatory News Service (RNS), news  agencies, brokers, stock market and individual investors, is a difficult  and time consuming task; At the beginning of the emergence of social  media, stock discussion forums pioneered an alternative source of  information for investors, specially retailers, supplementing the  traditional news media. A plethora of academic works study the  capability of these highly-specialized social channels to predict  returns and to detect abnormal behaviours in the stock market. \cite{Antweiler2004}  \cite{Campbell2011}  \cite{Sabherwal2011} \cite{Delort2012}. As online social media invaded the habits of  people, also, companies, brokers and other key roles in the financial  market began to share more and more factual information and informed  opinion in social media. The task of a financial analysts becomes more  challenging if we consider that not only have they to take decisions in  seconds but also deal with an increasing number of data sources  comprising of a continuous range of different features and quality.  Although social media may be a vehicle to fight against asymmetric  information in the financial market \cite{Billett2016}, the volume,  velocity and variety of its data make investing a heroic task,  especially for individual investors.

Given today, the co-existence of social media and traditional news media  (newspaper or online news media), the authors in \cite{Hu2016} compare the performance of social media and news media in  predicting returns in the Australian market. It is clear that social  media outperforms traditional media on the capability for readers to  interact or further spread information. Although the study shows a  significant effect of sentiments on one online discussion forum focused  on the Australian financial market, it found that the effect of new  media wasn't significant. The most interesting finding was that  sentiments on online investor forums do not seem to mirror the ones from  news media.There is in fact, an ongoing debate about to what extent  social media analytics would be the new predictor for the future returns  of stock-exchange traded financial assets.~We may even wonder if social  media can lead the direction of the financial market at some point in  the future as some authors suggested \cite{Zheludev2015}.

Although organizations have always used various corporate disclosure  channels to communicate directly and indirectly with investors, they are  now relying on social media to provide up-to-date information to  investors. {As the heart of publicly accessible Social Media, Twitter  has become a vital source for open source intelligence about natural  disasters, terrorist attacks, public health, politics, etc. Also,  Twitter is one of the most currently used platforms to share financial  information from companies, brokers, news agencies or individual  investors. Some recent works in the literature study this changes in  reporting financial news \cite{Xiong2016} \cite{Xiong2015} \cite{Miller2015}. As Twitter usage in this  context is definitively increasing; it is important to stress that,  according to \cite{Sprenger2014}, stock  microblogs exhibit three distinct characteristics about stock message  boards: (1) Twitter's public timeline may capture the natural market  conversation more accurately and reflect up to date developments;  (2) Twitter reflects a more ticker‐like live conversation which  allows micro-bloggers to be exposed to the most recent information of  all stocks and does not require users to actively enter the forum for a  particular stock; and (3) micro-bloggers have a strong incentive to  publish valuable information to maintain reputation (increase mentions,  the rate of retweets, and their followership), meanwhile financial  bloggers can be indifferent to their reputation in the forum. Providing  sensing, harvesting and analysing methods and tools of such information  could be very useful for many stakeholders such as businesses and  individuals making decisions to invest, stock market analysts and law  enforcement agencies.

All the roles on the financial market (investors, specialized news  agencies, etc.) are using Twitter to continually monitor the pulse of  the market and make decisions. The literature pays particular attention  to several ways in which the different market agents and participants  may use Twitter analytics. Taking the plethora of works related to  opinion mining on Twitter, one of the most researched areas is examining  consumer behaviour for financial purposes. Although real-time decisions  in the stock market are the most obvious manifestation of investment,  long-term investments are more related with consumer analysis, so that,  discovering the driving actors for sales and earnings is an active area  of research  \cite{Agarwal2017}. Secondly, the analysis of the relationships between Twitter behaviour  and stock share price is also a prime example of the increasing flow of  information between financial and Twitter universes. For instance, \cite{Ranco2015} investigated a 15-month period of Twitter data  including sentiment, concerning 30 stock companies registered on the Dow  Jones Industrial Average (DJIA) index. This work gave some insights  about sentiment and abnormal returns during the peaks of Twitter volume.  Specifically, the authors show that not only is there a strong  interaction between Twitter and the financial market in some moments  identified as ``known'' relevant events (i.e. quarterly announcements),  but similar results was observed in peaks not corresponding to any  expected news about the stock market. Also, \cite{Liu2015} used Twitter to identify and predict stock co-movement according  to firm-specific social media metrics and \cite{Shutes2016} studied tweets related to US market as indicator  of some (potentially new) information in the stock market rather than on  evaluating the problem of causality on stock prices.The results in \cite{Shutes2016} show that nearly a third of  the tweets in the study is associated with abnormal price movements so  that the authors suggest that Twitter is not a replacement for  traditional sources in financial market. For instance, Twitter lacks the  concrete trading recommendations which are common in other financial  information sources.

Consequently, there is a general consensus of the good sensing and  novelty characteristics of Twitter as a source of information for the  complex financial market. However, the provenance of contributions,  their different levels credibility and quality and even the purpose or  intention behind them makes Twitter not reliable enough as single source  for decision taking. }Our medium-term objective is a collaboration  project among the University of Vigo and the Manchester Metropolitan  University to deploy an architecture for real-time monitoring of  irregularities in the stock market. That architecture will apply data  mining and fusion technologies from a pool of social media feeds related  to the stock markets. In order to design the architecture, the  permeability of the different feeders should be analysed, that means, to  what extent a specific financial information feeder is permeable to  fraudulent and common irregularities in the financial market. The aim of  this paper is to analyse that permeability for the case of the  microblogging platform Twitter. The Intelligent System Group at the  Manchester Metropolitan University has worked in the detection of  irregularities form Financial Discussion Forums \cite{Owda2017}. Meanwhile, the Information \& Computing Lab of the University  of Vigo has a relevant know-how in applying Twitter analytics to a  variety of real life problems \cite{Khalifa2016} \cite{Servia2015}.

This paper analyses the permeability of Twittersphere, the total  universe of Twitter users and their habits, to a relevant event for the  financial market in the past few months: {the announcement about the  merge of Tesco PLC (hereinafter Tesco) and Booker Group PLC (hereinafter  Booker) to create the UK's Leading Food Business on the  27\textsuperscript{th} January 2017. Both companies, Tesco PLC and  Booking Group PLC are listed in the main market of LSE (London Stock  Exchange). }The findings show that the use of general purpose social  media data is permeable to financial-specific events. The work presented  is the first step in considering Twitter as a relevant feeder for taking  decisions regarding the financial market and detecting irregularities in  that market. An initial analysis of this experiment in \cite{Fernandez-Vilas2017} reported important insights when considering Twitter volume. This paper advances in the analysis of the Twitter permeability at that event by considering also the content of posts. Our hypothesis is that Twitter (although not a specific  financial forum) is permeable to financial events and this permeability  can be analysed by monitoring some specific features related to companies.

This paper is structured as follows. Section \ref{sec:2} introduces the Twitter efforts to accommodate financial information in a general-purpose  microblogging platform as well as related work in the area of the use of Twitter data for financial analysis. In Section \ref{sec:3}, we propose a model  of the position of Twitter in the financial universe as well as the path flows of information.{~ }Also, we introduce the experiment and a total  of five detailed research questions (RQ1 to RQ5) related to our hypothesis.The experimental methodology takes advantage of the extraction characteristics of Twitter APIs to construct a dataset specific for the merger (Section \ref{sec:4}). Then a series of investigations is conducted to address the five research questions and the impact is analysed according to its disturbance in terms of Twitter volume and features (RQ1, Section \ref{sec:5}), in terms of hashtag dynamics and topic modelling (RQ2, Section \ref{sec:6} and Section \ref{sec:7}), in terms of sentiment towards the actions (RQ3, Section \ref{sec:8}), in terms of geographical distribution of the posting (RQ4, Section \ref{sec:9}) and in terms of the rapidness \& synchronization (RQ5, Section \ref{sec:10}) with the LSE regarding the announcement and the stock share prices.{~ }Finally, Section \ref{sec:11} discusses our findings and introduces our ongoing work in the study of permeability of Twitter to financial events.

\section{Twitter \& the financial Market}
\label{sec:2}

It is fair to say that it was Twitter that popularised the term hashtag as well as its \# symbol to index keywords or topics so that people can  easily follow topics they are interested in. In 2012, Twitter unveiled a new clicking \& tracking feature for stock symbols known as Cashtags.  Cashtags are stock market symbols that can be included in tweets and when preceded with a dollar sign (for example \$VOD in regards to  Vodafone) become clickable.\cite{Hentschel2014} reported an exploratory analysis of public tweets in English, extracted via Twitter Firehose,  which contain at least one Cashtag from NASDAQ ({National Association of Securities Dealers Automated Quotation}) or NYSE (New York Stock  Exchange).{~ }The analysis concludes that the use of Cashtag is higher in the technologic sector, which seems to be related to the  technological profile of most of the Twitter users. In addition, the top 10 Twitter accounts according to the usage of Cashtags are companies or  news agencies which in the majority of cases correspond to automatic or semi-automatic Twitter accounts. This analysis also highlighted the  existence of relevant information behind the co-occurrence of Cashtags and the co-occurrence of Cashtags and Hashtags together. Other research  has examined the possible connections between Twitter information and financial market performance, that is the predictive value of  information gathered form social media to take decision about trading \cite{Ruiz2012} \cite{Sprenger2014}. Most of these works, based on the twitter data volume, also apply some sentiment analysis techniques in  order to distinguish the polarity of content and its{~ }impact on the financial market \cite{Oliverira2017}, \cite{Liew2016} \cite{Rajesh2016}  \cite{Cortez2016}.

Twitter is a valuable source of information even for the financial sector where contributors mainly fall into 5 categories \cite{Dredze2016}: (1) Journalists; (2) Companies and their representatives; (3) Government agencies; (4) Activist investors; and (5) citizen journalists (individuals). Also, the type of financial information is different, comprising not only of breaking news but rumours and speculations. According to \cite{Dredze2016}, this new financial media comes with new challenges as the huge volume of available data, the high number of repetition of the same information and, notably, a continuum of quality of the tweets, demands mechanisms measure credibility. Also \cite{Ceccarelli2016} suggests the importance of social media in the financial market, in particular Twitter, by analysing the popularity of Bloomberg tweets. Again, the authors agree on the fact that Twitter complements traditional news with speculations as well as off-the-cuff reporting, but also provides evidence that popularity within finance is not necessarily the same as popularity within other areas in Twitter, so that `novelty' seems to be a very impacting feature of the popularity of financial tweets. Also, the importance of the very first source of the news itself,{~ }the negative disclosures in this case, has been analysed in \cite{Elliott2017}. Their experiment found evidence that negative financial news influences investors' willingness to invest when the news comes from the Investor Relations Twitter account, but not when it comes from the CEO's Twitter account. Taking apart the concerns about information veracity and credibility,  freshness characteristics of Twitter also may have an important role in the field of High Frequency Trading (HFT) when traders make an investment position that is held only for very brief periods of time, even just seconds. At HFT, investors track social media to count for public behaviour and opinion to take their investor decisions so the the relationship between Twitter sentiment and  financial market instruments like volatility, trading volume, etc. and reports is investigated in \cite{Rao2014} with promising results in DJIA and NASDAQ-100.

But, Twitter is neither the only nor the biggest source of information about financial markets and, in this respect, {Figure 1} shows a reference model for the Twitter position regarding this universe of information and its relationships within the stock market. At any moment (time) and from anywhere (location), a variety of contributors in the financial sector (verified and unverified Twitter accounts) may post (tweet or retweet) pieces of information referring to a company by using either (1) the unique symbol \$cashtag, (2) \#hashtags related with the company for a specific purpose or related to a specific event and /or (3) simply the company name in its different forms (official, abbreviation, colloquial, etc.).{~ }At the same time that main agents in the financial market post into Twitter, the Twitter content maintains references to the financial universe outside via mentions, URLs, etc.{~ }Given apart the flows which characterise the permeability of the layer in between Twitter and the financial market, the information naturally spreads throughout the Twittersphere by the common re-tweeting and following mechanism in the platform. Even if the piece of news (post) comes from a human spread rumour or some cyber-attack which ends up being false information, \cite{doi:10.1177/0263276415583139} sustains about "the power of networked social and financial systems to connect autonomously and to produce a present without human oversight or governance". In our opinion, further actions \& research are needed to achieve a safer financial ecosystem for traders and investors.

\begin{figure}
\includegraphics[width=\textwidth]{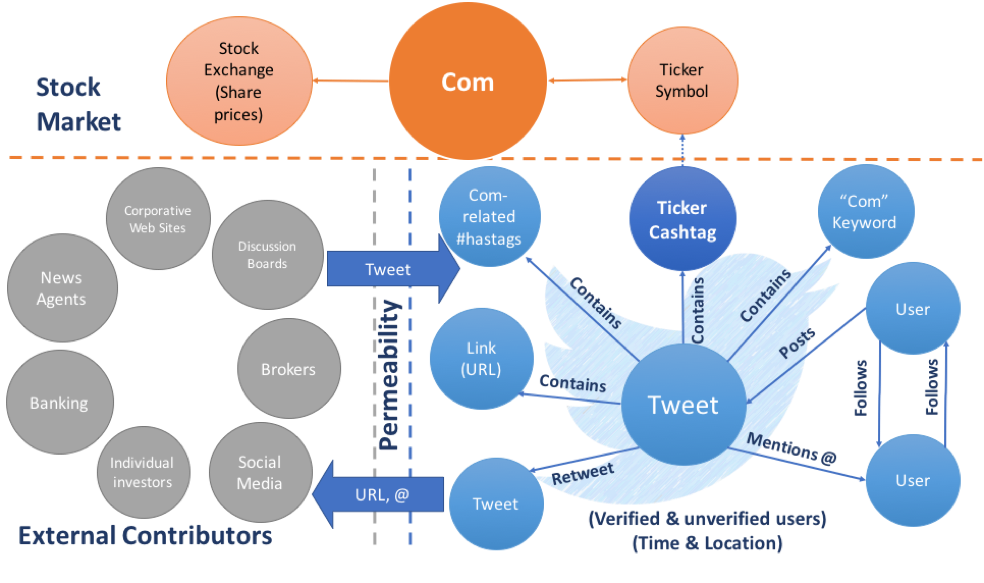}
\caption{\label{Figure1} Twitter position in the universe on financial information}
\end{figure}

\section{The experiment and the research questions}
\label{sec:3}

In this section, we describe the experimental methodology which will be used to answer our main hypotheses:{~}{ Twitter (although not a specific financial forum) is permeable to financial events and this permeability can be analysed by monitoring some specific features related to companies. To }address this high-level research question, we selected a random but relevant even, according to its impact on the financials spheres but also on the general public, at the beginning of year 2017. The most important feature for the selection of the event is the latter, that is, an event which being intrinsically financial, is of common interest for the general audience so that permeability can be measured beyond the spheres of financial experts in Twitter. 

The event under study, is {the }merge of Tesco and Booker on the 27\textsuperscript{th} January 2017, announced by the RNS (Regulatory News Service) of LSE (London Stock Exchange) at 7:00 a.m. GMT (Greenwich Mean Time). This action is modelled according to the reference model in Figure \ref{Figure1}, that is, we model Tesco on Twitter with the conceptual triplet (cashtag, hashtags, keyword), that is (\$TSCO, \#{[}Tesco{]}, ``Tesco'').{~ }This model represents the financial perspective of Tesco on Twitter (\$TSCO), specific comments about some Tesco issue on Twitter (\#{[}Tesco{]}: \#hashtag in some tweet with the keyword ``Tesco''), and general references to Tesco on Twitter (``Tesco'').{~ } As mentioned, our medium-term objective is a monitoring platform which continuously harvests signs scattered all over social media platforms to identify irregularities. Although in such an approach, not only Tesco, but Booker would be continuously monitored, the experiment faces the isolated behaviour of Tesco (one of the parts in the action) due to significance reasons, as the part with more presence in social media and due to design reasons, as the company being the focus of {real-time irregularity monitoring.}

Regarding permeability to financial events, {we may hypothesise that the permeability and the impact is not alike in the three perspectives which constitute the triplet.{~ }}The cashtag is invariably linked to financial news of a company but hashtags have a completely different dynamic. The hashtags related to a specific company will emerge and disappear dynamically according to the company decisions, marketing campaigns, consumer behaviour, etc. Tesco being a well-known company in the UK, the impact of financial events in hashtags might be limited and just visible in case of general-public financial events. Presumably, financial events should have a bigger impact on cashtag tweets (according to volume and context) than on hashtags (just altering their dynamics) or on topics of Tweets content. Nevertheless, this presumably different behaviour should be inspected.

Given our hypothesis: \textbf{\emph{``Twitter (although not a specific financial forum) is permeable to financial events and this permeability can be analysed by monitoring (1){~ }the name of companies as a keyword (``Tesco'' in this case), (2) the Cashtag of the company (\$TSCO) and (3) the hashtags related to that company.''}}

The experiment will attempt to answer the following research questions (RQ\emph{n}) related with the permeability of Twitter to financial events:
\begin{itemize}
\item  \textbf{RQ1: Is the event impact on \$cashtag-content greater than on   \#hashtag-content or on general content?}
\item  \textbf{RQ2: Although \$cashtag-content has a clear financial   orientation, can a financial load be also perceived in the \#hashtag   dynamics as in the general content?}
\item  \textbf{RQ3: Can we measure change in sentiment related to the   financial event through time?}
\item \textbf{RQ4: Is the event impact on Twitter dependent on the   location?}
\item \textbf{RQ5: How is the permeability in terms of rapidness in response   and synchronisation with the dynamics of LSE? that is, rapidness in   relation to the RNS announcements of LSE and synchronisation with stock share prices.}
\end{itemize}

On the basis of the proposed RQs, and even though we are reporting a single case study, we would like to highlight that our triplet approach can be applied to the general monitoring of financial event on Twitter disregarding the market and the sector if we consider announcement services and stock prices in other regulated financial markets.

\section{Twitter data mining \& the dataset}
\label{sec:4}

\begin{figure}
\includegraphics[width=\textwidth]{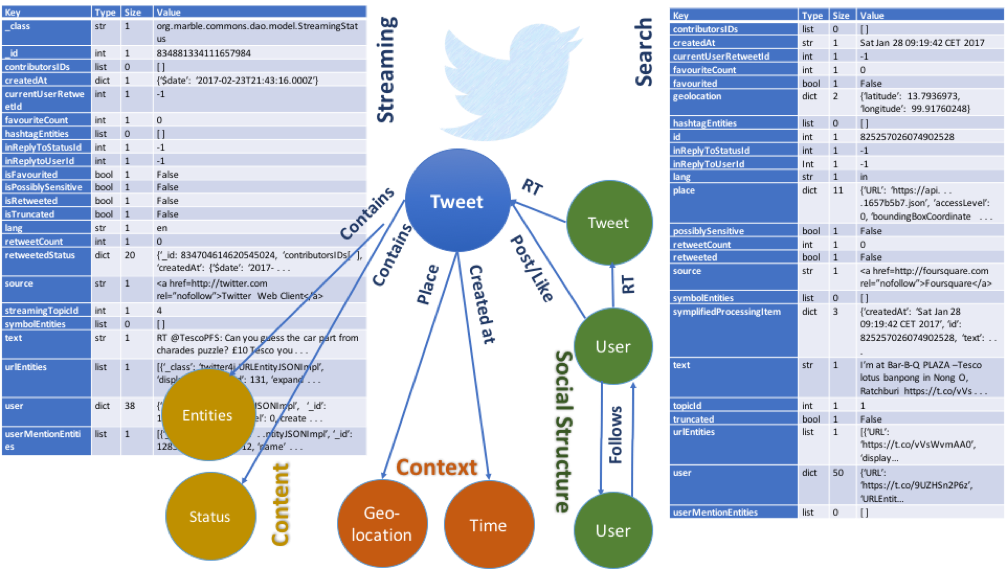}
\caption{\label{Figure2} Twitter APIs and a model of the Twittersphere.}
\end{figure}

There are three different ways to obtain Twitter data: Search API, Streaming API and Firehose. ~The Twitter Search API provides the endpoints to recover tweets that were published in the previous two weeks, with the possibility of filtering according to several criteria. On the other hand, Twitter Streaming API returns 1\% of the tweets that match some search parameters in real time. Finally, Twitter Firehose provide access to the 100\% of the tweets, but it is not a free-access API.{~ }Twitter APIs are constructed around four main ``objects'': \emph{Tweets}, \emph{Users}, \emph{Entities} (hashtags, URLs, mentions and media in a tweet) and \emph{Places}. From theses object and taking time as the Twitter's backbone, we consider a Twitter model which contemplates 3 orthogonal perspectives: The content, the social structure and their spatio-temporal context (see Figure \ref{Figure2}).

The anatomy of these objects is described in the Twitter Developer documentation. With regard to this work, as the experiment does not focus on the dynamic spreading of information on Twitter (via retweeting, mention or like mechanisms), we select the following features (existing in both APIs under different field names) for the analysis, all of them accessible from a Tweet object:

\begin{itemize}
\item
  Content perspective: the status update (\texttt{Tweet:text}) and the entities (\texttt{Tweet:entities}), specifically \texttt{hashtags} (including cashtags) and \texttt{urls}.
\item
  Context perspective: the post time of the status update   (\texttt{Tweet:created\_at}) and, if available, also the place  by feature \texttt{Tweet:coordinates} and feature \texttt{Tweet:place:bounding\_box}).
\item
  Social Perspective: User (\texttt{Tweet:user}, specifically the field   verified).
\end{itemize}

{There are some differences between the Searching API and the Streaming API illustrated in Figure \ref{Figure2}. One difference being the time direction -{~ }the most relevant one to our experiment. The Search API }goes back in time, whilst the streaming API goes forward. Moreover, there are other differences related to mainly the format and the rate limit rules. As it is shown in Figure \ref{Figure2}, the search and Streaming API does not return data in exactly the same format but the differences are really insignificant to data analysis. Regarding their extracting capacity, forums contain plenty of discussion about this issue which has not ever made enough clear from Twitter officially.

According to the proposed model (Figure \ref{Figure1}) the experiment will analyse the permeability of Twitter to financial events by the inspection of the triplet (\$TSCO, \#{[}tesco{]}, `tesco'), defined as:

\begin{itemize}

\item   \$TSCO, the set of tweets where the ticker symbol \$TSCO is an   entity;
\item  \#{[}tesco{]}, the set of tweets with at least one hashtag and   containing the keyword `tesco'; and `tesco',
\item  the set of tweets containing the keyword `tesco', no matter whether   they contain a hashtag or not. 
\end{itemize}

\begin{figure}
	\begin{tabular}{ll}
\includegraphics[width=0.45\textwidth]{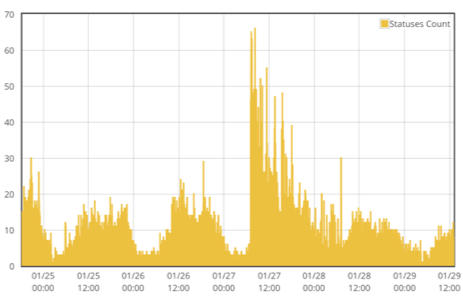} &
\includegraphics[width=0.45\textwidth]{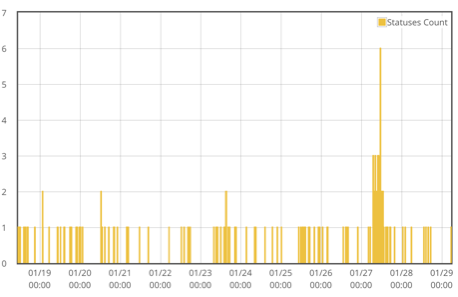}
\end{tabular}
\caption{\label{Figure3} Total Twitter volume for the keyword `tesco' (left) and for the cashtag \$TSCO (right) by merging (without duplicates) data returned from queries to Search API (backwards) and Steaming API (backwards).}
\end{figure}

The extraction strategy firstly used the Twitter Search API to recover the information backwards before the announcement on 27\textsuperscript{th} of January 7:00 a.m. GMT and the streaming API was used to recover information forwards until 27\textsuperscript{th} February (one month later).{~ }This data streaming collection, just after the announcement, was used to visualise the impact of the announcement and the time the behaviour of tweets concerning Tesco resumes a regular pattern again.{~ }The results of the combination of the search and streaming results is shown in Figure \ref{Figure3}.

 Once the behaviour becomes regular, we used the Search API again to obtain a regular dataset to compare with the Search API results recovered just after the merge action was announced. Clearly, the Twitter Search API is not appropriate for continuous analytical monitoring and as a data source to allow real-time decision making. It is not intended and does not fully support the repeated constant searches that would be required to deliver 100\% coverage. However, the experiment in this paper is limited to one individual company, 2 keywords and timelines in the scale of weeks. Therefore, the Search API provides a better coverage than the Streaming API (1\% according to the Twitter official information), if the superior filtering characteristics of the Search API are used. Nevertheless, as the Search API has a limit on the number of tweets recovered, to get the whole data during the period under study (see Table \ref{Table1}), we repeatedly ask Twitter for the most recent results backwards by windowing the searches according to the publication date and merging results according to the post Id. In this way, we guarantee a fair comparison according to the volume of data since, in any manner, we should compare the Search API with Streaming API results. According to that, and to give response to the research questions, we use the Search API queries to cover the time periods defined in Table \ref{Table1}. Mainly, the 4 periods in this table capture the Twitter volume retrieved by the Search API during the very same days of the week, Wednesday to Sunday, around the announcement on Friday 27\textsuperscript{th} of the merge (pre-announcement and post-announcement periods) and when behaviour becomes regular at the corresponding week days. Two types of Search API queries were considered, (1) query with the term `tesco' for the Tweet:entities and the {Tweet:text}{~ }entities, to capture post related with Tesco and post also containing a hashtag related with Tesco ( a total of 70,793 tweets) and (2) query with the term `\$TSCO' for Tweet:entities (a total of 151 tweets). It is fair to mention that `tesco' volume is several orders of magnitude higher than `\$TSCO' given the high visibility of the company Tesco in social media as a vehicle for marketing and consumer engagement.

\begin{table}[]
\centering
\caption{\label{Table1} Time Periods (GEM Time) in the experiment (extracted with Twitter Search API)}
{\small
\begin{tabular}{l|l|l|ll|}
\hline
\rowcolor[HTML]{9B9B9B} 
\multicolumn{1}{|c|}{\cellcolor[HTML]{9B9B9B}\textbf{Name/Period}} & \multicolumn{2}{l|}{\cellcolor[HTML]{9B9B9B}\textbf{‘tesco’}}              & \multicolumn{2}{l|}{\cellcolor[HTML]{9B9B9B}\textbf{\$TSCO}}                          \\
\rowcolor[HTML]{9B9B9B} 
                                                                   & \multicolumn{2}{l|}{\cellcolor[HTML]{9B9B9B}\textbf{incl. \#{[}tesco{]}}} &                                                             &                         \\ \cline{2-5} 
\rowcolor[HTML]{9B9B9B} 
\multicolumn{1}{|l|}{\cellcolor[HTML]{9B9B9B}}                     & \textbf{Total}                       & \textbf{Per/hour}                   & \multicolumn{1}{l|}{\cellcolor[HTML]{9B9B9B}\textbf{Total}} & \textbf{Per/hour}       \\ \hline
Pre-announcement                                                   &                                      &                                     & \multicolumn{1}{l|}{}                                       &                         \\
25th Jan 00:00- 27th Jan 07:00                                     & \multirow{-2}{*}{11,817}             & \multirow{-2}{*}{214.85}            & \multicolumn{1}{l|}{\multirow{-2}{*}{12}}                   & \multirow{-2}{*}{0.218} \\ \hline
Post-announcement                                                  &                                      &                                     & \multicolumn{1}{l|}{}                                       &                         \\
27th Jan 07:00 - 29th Jan 23:59                                    & \multirow{-2}{*}{25,547}             & \multirow{-2}{*}{393.03}            & \multicolumn{1}{l|}{\multirow{-2}{*}{91}}                   & \multirow{-2}{*}{1.4}   \\ \hline
Regular 2-weeks-after                                              &                                      &                                     & \multicolumn{1}{l|}{}                                       &                         \\
8th Feb 00:00-10th Feb 07:00                                       & \multirow{-2}{*}{13,417}             & \multirow{-2}{*}{243.94}            & \multicolumn{1}{l|}{\multirow{-2}{*}{26}}                   & \multirow{-2}{*}{0.473} \\ \hline
Regular 2-weeks-after                                              &                                      &                                     & \multicolumn{1}{l|}{}                                       &                         \\
10th Feb 07:00 - 12th Feb 23:59                                    & \multirow{-2}{*}{20,012}             & \multirow{-2}{*}{307.88}            & \multicolumn{1}{l|}{\multirow{-2}{*}{22}}                   & \multirow{-2}{*}{0.338} \\ \hline
\end{tabular}}
\end{table}

\section{Impact on Twitter volume}
\label{sec:5}

In this section, we detail the impact of the event by analysing the variation in the number of tweets (volume) so that a quantitative measure of Twitter permeability to a financial event can observed. During this part of the analysis some irregularities were discovered which related to an inconsistency in the named scheme of tickers in Twitter. In particular, to our knowledge, Twitter has not promoted the specific distinction among financial markets so that the uniqueness of ticker symbols inside a market disappear in the Twittersphere. That is the case of \$TSCO cashtag which corresponds to Tesco PLC in the LSE and to \emph{Tractor Supply Company} in the NASDAQ, the second stock exchange in USA. So, the returned results to a \$TSCO query included tweets related to Tesco PCL and also to Tractor Supply Company. If the cashtags is the entity to aggregate information around a specific company and, consequently, to allow the spreading of such information, some kind of market prefix should be used, especially in the times when companies are increasingly global.

\begin{figure}
	\begin{tabular}{c}
\includegraphics[width=\textwidth]{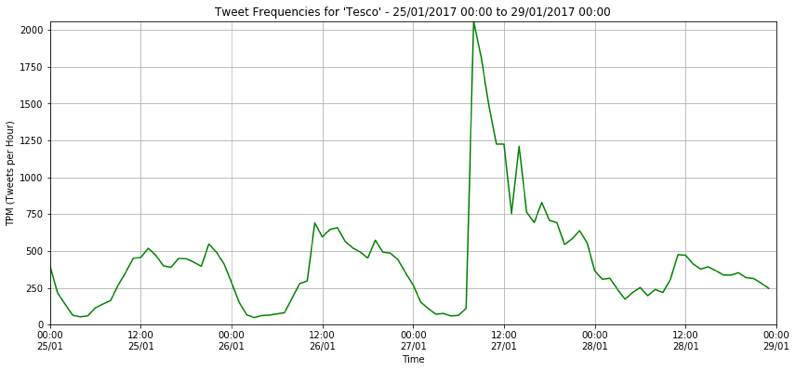}\\
\includegraphics[width=\textwidth]{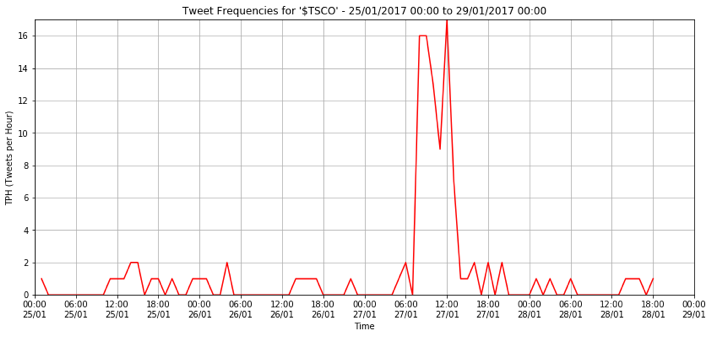}
\end{tabular}
\caption{\label{Figure4} Time series of the `tesco' and \$TSCO dataset from 25th January to 29th January.}
\end{figure}

Figure \ref{Figure4} shows the temporal series in tweets per hour (TPH) scale.{~ }Although it is quite obvious that the number of TPH in the `tesco' dataset is up several orders of magnitude higher than those of the \$TSCO dataset, the peak behaviour is more acute in the \$TSCO one. As it is shown in Table \ref{Table2}, considers the hourly volume of the `tesco' dataset on the 27\textsuperscript{th} January. There are no outliers during the day, with the peak value of 2,057 tweets occurring in the sample from 8:00 to 9:00. Nevertheless, there are 3 outliers in the \$TSCO dataset: samples 8:00-9:00, 9:00-10:00, corresponding to the time just after the announcement and 12:00-13:00, being consistent with previous studies about social timing in \cite{Adnan2014}, which showed peak activity during lunch time in different cities around the world.

\begin{table}
\caption{\label{Table2} Peak behaviour on the 27\textsuperscript{th} January for `tesco' and \$TSCO}
\begin{tabular}{c}
\toprule \\
\includegraphics[width=\textwidth]{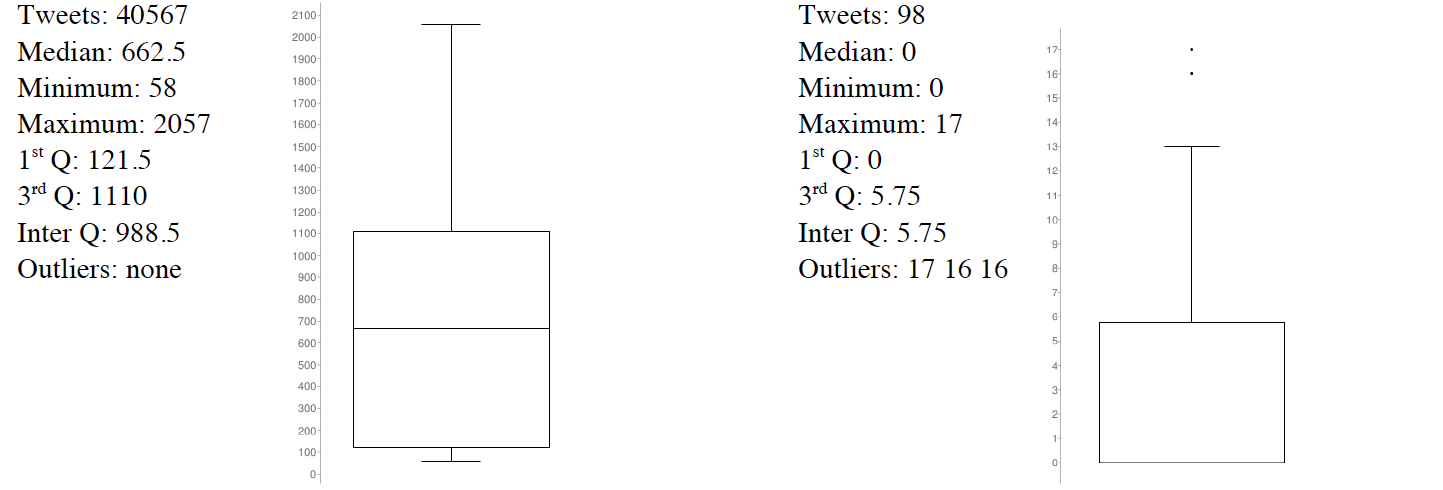} \\
\bottomrule
\end{tabular}

\end{table}

Apart from the peak comparison, we inspect the disturbance on other dataset features before and after the event, also comparing these dates with the regular behaviour 2 weeks later. In Table \ref{Table3}, we compare the behaviour of the main features of `tesco' data (in green) and \$TSCO' data (in blue).{~}

\begin{table}
\caption{\label{Table3} Variability of features in `tesco' and \$TSCO data before and after the announcement in terms of regular behaviour.}
\includegraphics[width=\textwidth]{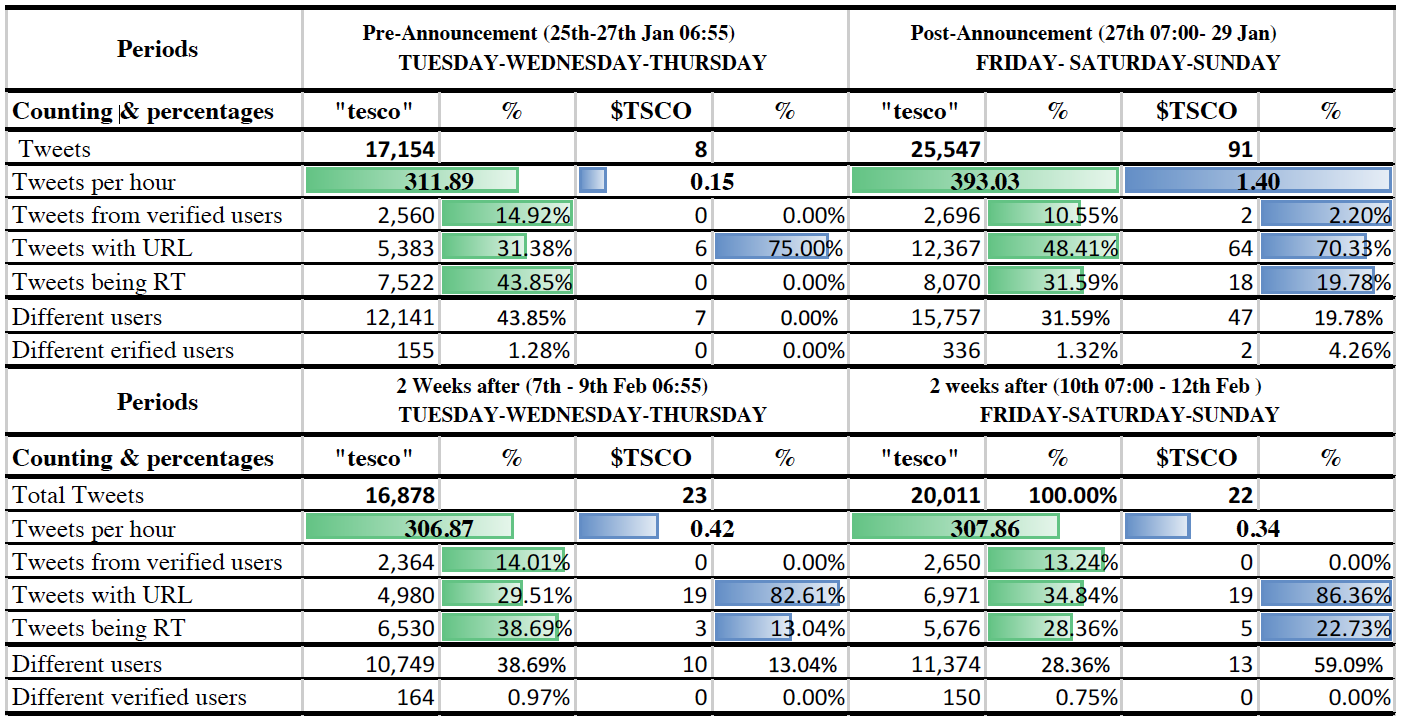}
\end{table}

Firstly, we can observe the increase of Tweets per hour during the post-announcement, compared to the regular period,{~ }is more acute in the \$TSCO data (1.40 vs 0.34) than in the `tesco' data (393,03 vs 307.86).Secondly, the percentage of tweets which contain a URL are significantly higher in the \$TSCO data (from 75\% to 82.61 \% in the different periods) with respect to the `tesco' one (from 29.51\% to 48.41\%), which is a result of the professional and financial orientation of the \$TSCO data as a channel to spread facts and news rather than opinions and sentiments. Finally, the retweeting activity is higher in the announcement periods (pre- and post- with 43.85\% and 31.59\%) compared to the regular periods (38.69\% and 28.36\%) for the `tesco' data. The increase of retweeting is, by nature, linked to the need or desire of spreading a piece of content but, the reason behind may be different as, in fact, it is in our case study:{~ }retweeting the `tesco' keyword is mainly related with a Tesco campaign involving retweeting (see section VI for the details) meanwhile retweeting in \$TSCO data is mainly linked to spreading the information about the merge (post-announcement) and about other financial news related with Tesco PLC. The information about this retweeting activity is expanded upon in the following sections of this paper. Finally, we observe the invariability on the number of verified users either along all the periods and along the `tesco' and \$TSCO data. 

Both the peak behaviour (Table \ref{Table2}) and the feature comparison (Table \ref{Table3}) provide an answer to \textbf{RQ1}, that is, the {\textbf{event impact on \$cashtag-content{~ }is greater than on \#hashtag-content or on general content.}}

\section{Impact on Hashtags Dynamics}
\label{sec:6}

{Given the volume of Tweets, the spreading of the event throughput the Twittersphere should presumably have two effects: the change on the hashtag dynamics and the change on the main topics in tweets. More than this effect, what we would like to }measure is to what extent this disturbance can be perceived differently in `tesco' and \$TSCO data. Presumably, and contrarily to the volume analysis, the content impact is expected to be more marked or noticeable in the `tesco' data. So that, financial topics, which are the core of cash-tagged tweets, would conquer the whole Tesco content as a result of the good permeability to this financial event in Twitter, should be studied.

Focusing on the analysis on hashtags as a measure of ephemeral or nor so ephemeral interest, Table \ref{Table4} shows the disturbance of the event in hashtag dynamics in the `tesco' data, which is considered data related with the company without any specific financial bias. In terms of the 5 most frequent hashtags during the periods considered in the experiment, we clearly perceive that, only during the period Post-Announcement, financial hashtags emerge:{~ }\#teschoshareprice and \#booker. We consider \#booker a hashtag related with the action given its co-occurrence with the keyword `tesco', the reverse may be not true.{~ }Also, we discover a persistent hashtag during all the periods: \#essothursdays, related with the equally named promotion campaign which allows Twitter and Facebook users to obtain £100 Tesco voucher provided that they follow the Twitter account @GB\_Esso and retweet the competition tweet.

\begin{table}
\caption{\label{Table4} Disturbance on Hashtag Dynamics for the `tesco' dataset.}
\includegraphics[width=\textwidth]{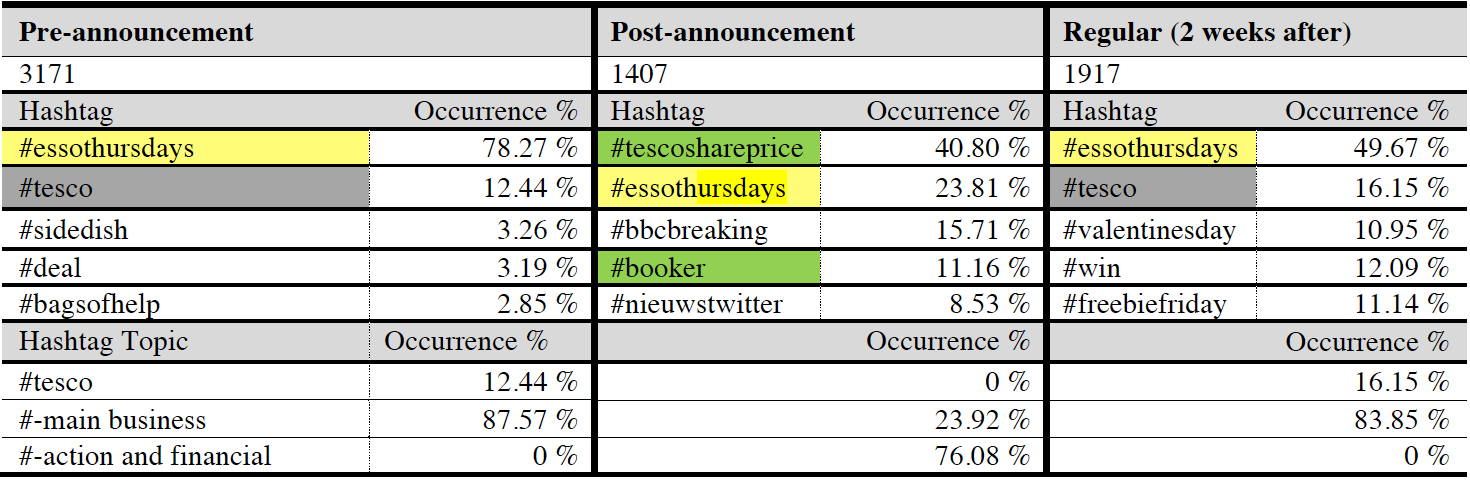}
\end{table}

\begin{figure}
\includegraphics[width=\textwidth]{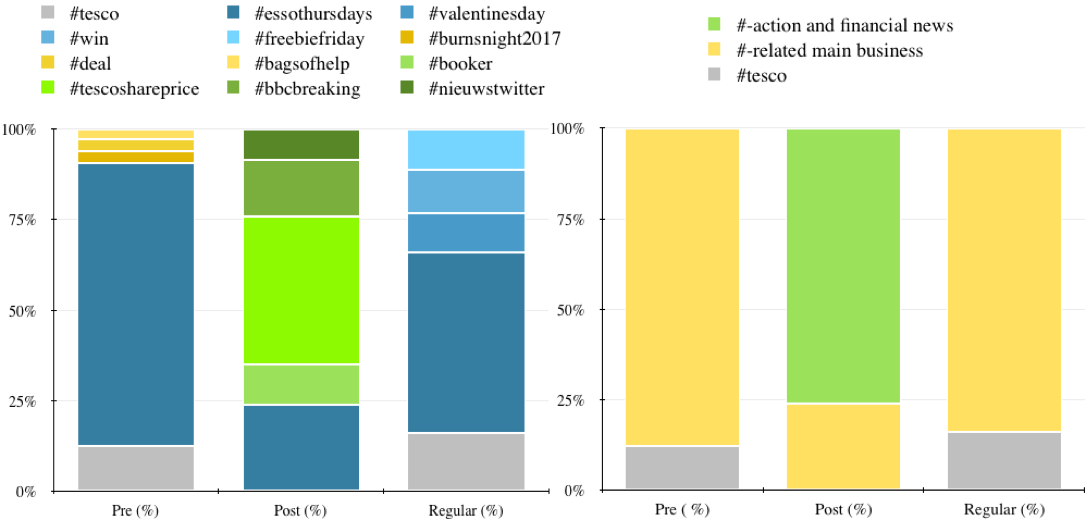}
\caption{\label{Figure5} Disturbance on Hashtag Dynamics for the `tesco' dataset (left) and aggregation of the hashtags in semantic families.}
\end{figure}

Regarding hashtag disturbance at a more general level, the bottom part of Table \ref{Table4} shows the percentage of occurrence of the main hashtag of the company (\#tesco), hashtags related with main business of the company (\#-main business), and hashtags related with merge and financial aspects (\#-action and financial). Also, Figure \ref{Figure5} shows hashtags in terms of semantic families. The left chart shows the main topics reported in the `tesco' data and the right chart distinguishes the three main categories within these topics: ``Tesco'' in grey, ``Main Business'' in yellow and ``Action and Financial news'' in green).{~ }It is clear that, from the point of view of the hashtags dynamics, the merge reporting achieves a high position in the `tesco' dataset after the announcement. During the pre-announcement period, hashtags (apart for the generic one \emph{\#tesco} in grey) are related to marketing campaigns (yellow in the aggregated chart on the right). It is only during the post-announcement period that financial-related hashtags emerge: \emph{\#tescoshareprice} being 40.80\% and \#booker being the 11.16\% of the top hashtags. Also, it is remarkable that the emergence of the hashtags related to the breaking news \emph{(\#bbcbreaking} and \#nieuwstwitter) and that even \#tesco disappears in the top-5 hashtags during the post-announcement period (Table \ref{Table4}). All of financial-related and breaking news hashtags completely disappear in the period we consider as regular in terms of Tesco experiment (see also Table \ref{Table4})

Lastly, regarding the analysis of hashtags in the \$TSCO dataset, it is not strange than during the regular period none of the tweets in the dataset contains a hashtag, so no extra specific information to follow and spread information is provided, apart from the pseudo-hashtag \$TSCO. However, during the post-announcement period 15,38 \% of the tweets in the \$TSCO dataset contains at least one hashtag. Nevertheless, the number of tweets during this period is 91 of which 16 include at least one hashtag. Although we prefer not to analyse the hashtag dynamics with this small number, we can appreciate a disturbance with regards to the regular period.{~ }This result talks about the malfunction of \#hashtags in the \$TSCO dataset. If \#hashtags are the Twitter tool to aggregate semantically-related content and spread it, this feature is almost absent in the \$TSCO dataset. We cannot perceive a hashtag disturbance related to the action, because in the financial island of Twitter the \$cashtag is the main Twitter tool to aggregate and spread content. What we can only appreciate with respect to the regular behaviour is the increase on the number of hashtags which accompany the pseudo-cashtag \$TSCO.

According to these results we can establish the fulfilment of \textbf{RQ2, in that Twitter permeability to the merge can be perceived not only on \$cashtag-content, which has a clear financial orientation, but also in the \#hashtag dynamics.}

\section{Impact on the Vocabulary \& Topics}
\label{sec:7}

When analysing the impact of the action on the Twitter content, the complete response to \textbf{RQ2 (}{\textbf{can a financial load be also perceived in the \#hashtag dynamics as in the general content?)}} \textbf{} should be studied also from the point of view of the main topics people post about Tesco. For that, we should extract these topics from the tweets content in `tesco' data and compare them with the ones in the specifically financial \$TSCO data. Unfortunately, the number of tweets in the \$TSCO dataset is relatively low. According to this, we decided to automatically split the `tesco' and \$TSCO data according to an automatic financial annotation by using a simple vocabulary related with general financial terms and specific terms related with the action.

The first RNS related to the event was published 27\textsuperscript{th} January 2017 as RNS number: 2907V under the title ``TESCO \& BOOKER ANNOUNCE MERGER''. The body of the announcement contains the following preface ``The boards of Tesco PLC (''Tesco``), the UK's leading food retailer, and Booker Group plc (''Booker``), the UK's leading food wholesaler, are pleased to announce that they have reached an agreement on the terms of a recommended share and cash merger (the ''Merger``) to create the UK's leading food business.'' To measure the financial burden of the topics in tweets content and the disturbance of this burden before and after the announcement, we apply a priori knowledge by selecting a simple vocabulary by inspecting the common terms in the content of the \$TSCO dataset and the content of the RNS. Two categories of terms were created:

\begin{itemize}
\item Generic financial terms: Board of Directors, Executive Manager, CEO,   Chairman, Director, Share, shareholder, revenue, return, capital,   investment, dividend, Mix \& Match.
\end{itemize}

\begin{itemize}
\item Specific action terms: Merge, take over, announcement, RNS, Merger,   Tesco, Booker, Combined Group, Stewart Gilliland (extracted for the   RNS announcement and the main actors in the action. The announcement is 85 pages long).
\end{itemize}

Regarding the representativeness of our vocabulary, the frequency of terms in the `tesco' data was 7.11\% and is much lower than the frequency in the \$TSCO data which is 47.68\%. We remark that these percentages referred to the total corpus (content of all the tweets in the dataset), not the total number of tweets in the datasets. Regarding the number of tweets containing at least one term in the vocabulary, we observe the following results in `tesco' dataset:{~ }{2.07\%~(245 of 11817) before the announcement and 9.52\% (2437 of 25547) after the announcement. This difference is obviously not so big in the \$TSCO dataset: 25.00\% (3 of 12) before the announcement and 27.47\% (25 of 91) after the announcement (the increasing can be explained because of the inclusion of terms specifically related with the action). }These differences in percentages is what these datasets should presumably exhibit according to the clear financial orientation of \$TSCO as a pseudo-hashtag for the financial facts related with Tesco PLC. So that, our simple vocabulary allows us to measure the financial load of a set of tweets.

According to the vocabulary, we split the `tesco' dataset (pre- and post-announcement) into a set of tweets containing at least one term in the vocabulary, hereinafter referred as \emph{`tesco' financially-oriented subset}; and the set of tweets not containing any term in our vocabulary, hereinafter referred as \emph{`tesco' general subset}. Then, for uncovering the main topics, we adopted a simple bag-of-words model \cite{Wu2010}. Under this approach tweet words in the status content are only considered according to their relative frequency and not according to their order within the document. In the bag-of-word result, we consider the proper cleaning precautions (i) by lowercasing and removing stop words; and (2) by specifically considering Twitter jargon (i.e. HT, QOTP). Finally, terms are contextualised according to the tweet semantics and categorised into topics. As a result, we obtain the contextualised \& categorised occurrence of terms in {Table \ref{Table5}, and the evolution of the size of the considered categories on the `tesco' corpus in Figure \ref{Figure6}.

\begin{table}
\caption{\label{Table5} Term occurrences in subsets of the `tesco' dataset (General and Financially-oriented)}
\includegraphics[width=\textwidth]{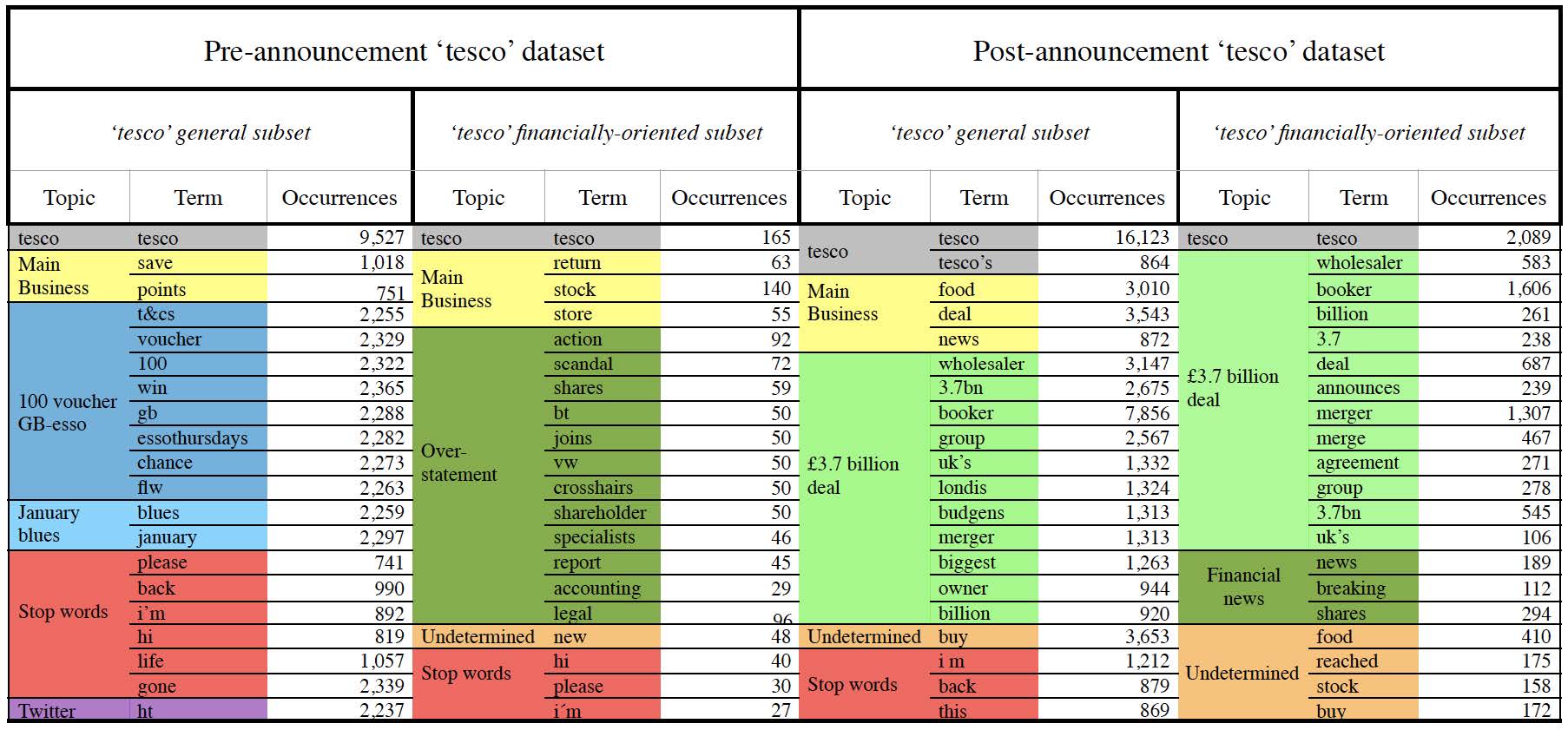}
\end{table}

More specifically, Table \ref{Table5} contains the contextualisation of the 20-most frequent terms inside the complete content of the tweet according to an n-gram strategy and manual annotation. We can obviate the term `tesco' in its different forms (tesco, tesco's in grey) from the analysis, since the dataset has been obtained from different `tesco' search queries to Twitter. Regarding the general annotated subset of{~ }the `tesco' dataset, before 27\textsuperscript{th} 7:00 a.m. all most frequent terms are either stop words (including Twitter jargon) and terms related with the main Tesco PLC business. Even though, we can distinguish a set of terms related with a specific Tesco campaign (the one previously mentioned in the appearance of \#essothursdays in blue) and also a topic `January blues' which can be considered seasonal in the UK, even in the context of a dataset specific for the company Tesco PLC. Just after the announcement, the scene totally changed and neither the campaign nor the seasonal topic appear but they are replaced by terms related to the Tesco PLC merge with Booker. Therefore, topic discovering exhibits the same kind of disturbance than hashtags, that is, the impact of the merge can be perceived in the \emph{`tesco' general subset} after the announcement. After this analysis, we can establish the fulfilment of the complete \textbf{RQ2\emph{:}} {\textbf{Given that \$cashtag-content has a clear financial orientation, this financial burden can also be perceived in the \#hashtag dynamics and in the general content.} So that, Twitter is highly permeable to the financial event considered in this experiment.}

\begin{figure}
\includegraphics[width=\textwidth]{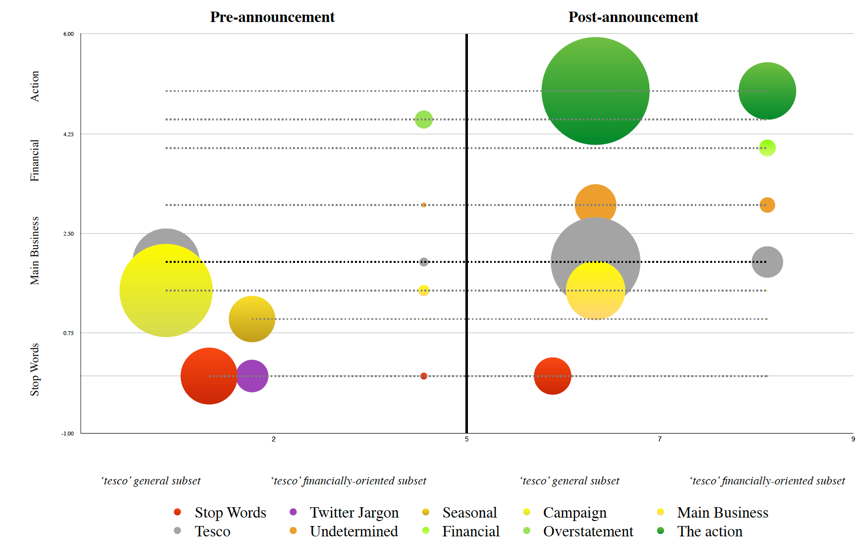}
\caption{\label{Figure6} Evolution of Main topics in the `tesco' corpus.}
\end{figure}

Regarding the \emph{financially-oriented subset} of `tesco', the most frequent terms are obviously related with Tesco PLC financial perspective before and after the announcement but, once again, the scene changes totally from one day to another. With the financial lens the filtering vocabulary provides, a topic emerges referring to another relevant financial event related with Tesco PLC. On Jan 24\textsuperscript{th}, a Tesco spokesman announced that the firm was facing a claim for damages about profit overstatement. This topic is, therefore, relevant enough to emerge in the financial subset but is disguised by the contents related with the main business in the general subset. What we have is a financial event, in the same company, which does not surpass the imaginary threshold of permeability to the general Twittersphere. This can be clearly perceived in the Figure \ref{Figure6} where bubbles are coloured according to the main topics (green scale for financial and yellow scale for main business, grey for Tesco in the middle of these two worlds) and sized according to the number of term occurrences. A big green bubble appears in the general post-announcement subset but, in the financially-oriented subset, the green bubble related to the overstatement is replaced by the green bubble related to the merge.

\section{Impact on polarity}
\label{sec:8}

{Unlike topic modelling, sentiment analysis is about applying natural language processing (NLP) to mine the subjective impression beyond the actual facts and to measure the individual sentiment or reactions toward certain products, people or ideas by revealing the contextual polarity of the information. The python library \emph{TextBlob} \cite{Loria2014} was used to discern }how positive or negative the sentiment in the tweet is. It is worth stating at the outset that we do the analysis with the same polarity classifier in two quite different datasets, the more general `tesco' data, which contains plenty of sentiments, and the financial \$TSCO dataset under the lens of the cashtag, more oriented to facts or as much opinions than sentiment behind human beings.{~ }Also, we used a polarity classifier previously trained on general Tweets and not specifically trained for financial content and its peculiar jargon. With all these precautions in mind, and taking into account that we are facing a single event, we provide an interpretation of the sentiment analysis (Table \ref{Table6}) supported by the knowledge acquired throughout the previous sections.

\begin{table}[]
\centering
\caption{\label{Table6} Sentiment Analysis results for the `tesco' and \$TSCO dataset}
{\small
\begin{tabular}{lllllllll}
                               & \multicolumn{4}{l}{\cellcolor[HTML]{C0C0C0}\textbf{{\tiny Pre-announcement (25-27 Jan 06:59 am)}}}                              & \multicolumn{4}{l}{\cellcolor[HTML]{C0C0C0}{\tiny}\textbf{{\tiny Announcement (27 07:00am-29 Jan)}}}                                   \\
                               & \multicolumn{4}{l}{\cellcolor[HTML]{C0C0C0}{\tiny TUESDAY-WEDNESDAY-THURSDAY}}                                                       & \multicolumn{4}{l}{\cellcolor[HTML]{C0C0C0}{\tiny FRIDAY-SATURDAY-SUNDAY}}                                                           \\ \cline{2-9} 
\multicolumn{1}{l|}{}          & \multicolumn{2}{l|}{\cellcolor[HTML]{C0C0C0}\textbf{‘tesco’}} & \multicolumn{2}{l|}{\cellcolor[HTML]{C0C0C0}\textbf{\$TSCO}} & \multicolumn{2}{l|}{\cellcolor[HTML]{C0C0C0}\textbf{‘tesco’}} & \multicolumn{2}{l|}{\cellcolor[HTML]{C0C0C0}\textbf{\$TSCO}} \\ \cline{2-9} 
\multicolumn{1}{l|}{}          & \multicolumn{1}{l|}{Tweets}   & \multicolumn{1}{l|}{\%}       & \multicolumn{1}{l|}{Tweets}  & \multicolumn{1}{l|}{\%}       & \multicolumn{1}{l|}{Tweets}   & \multicolumn{1}{l|}{\%}       & \multicolumn{1}{l|}{Tweets}  & \multicolumn{1}{l|}{\%}       \\ \hline
\multicolumn{1}{|l|}{Positive} & \multicolumn{1}{l|}{181}      & \multicolumn{1}{l|}{55.20\%}  & \multicolumn{1}{l|}{1}       & \multicolumn{1}{l|}{12.50\%}  & \multicolumn{1}{l|}{599}      & \multicolumn{1}{l|}{24.60\%}  & \multicolumn{1}{l|}{30}      & \multicolumn{1}{l|}{33.00\%}  \\ \hline
\multicolumn{1}{|l|}{Neutral}  & \multicolumn{1}{l|}{90}       & \multicolumn{1}{l|}{27.40\%}  & \multicolumn{1}{l|}{7}       & \multicolumn{1}{l|}{87.50\%}  & \multicolumn{1}{l|}{1636}     & \multicolumn{1}{l|}{67.10\%}  & \multicolumn{1}{l|}{43}      & \multicolumn{1}{l|}{47.30\%}  \\ \hline
\multicolumn{1}{|l|}{Negative} & \multicolumn{1}{l|}{57}       & \multicolumn{1}{l|}{17.40\%}  & \multicolumn{1}{l|}{0}       & \multicolumn{1}{l|}{0.00\%}   & \multicolumn{1}{l|}{202}      & \multicolumn{1}{l|}{8.30\%}   & \multicolumn{1}{l|}{18}      & \multicolumn{1}{l|}{19.80\%}  \\ \hline
\multicolumn{1}{l|}{Total}     & \multicolumn{1}{l|}{328}      & \multicolumn{1}{l|}{}         & \multicolumn{1}{l|}{8}       & \multicolumn{1}{l|}{}         & \multicolumn{1}{l|}{2437}     & \multicolumn{1}{l|}{}         & \multicolumn{1}{l|}{91}      & \multicolumn{1}{l|}{}         \\ \cline{2-9} 
\end{tabular}}
\end{table}

Regarding the `tesco' dataset, the analysis during the regular period in the experiment exhibits a 33.19\% positive, 44.54\% neutral and 22.27\% negative sentiment load. Contrarily, just before the announcement the same dataset increases its positive sentiment to 55.2\% which seems to be a result of the GB Esso promotional campaign to obtain £100 Tesco Voucher. Although this promotional campaign continues much time after the announcement, the sentiment in the `tesco' dataset turned into more neutral, 67.1\%, due to the penetration of the news about the Tesco \& Booker merge to the general Twitter audience. Given that this information is for the general public more factual than sentimental, neutrality increases just before the announcement. The analysis of the \$TSCO dataset is the reverse way, tweets under the \$TSCO umbrella are usually factual or neutral, or at least neutral for a general polarity classifier, but when an impacting financial event occurs neutrality vanishes and a bi-polarity emerges in response to the merge. Some example tweets and their corresponding polarity are provided for illustrative purposes only in Table \ref{Table7}. To conclude, and with the precautions mentioned, we can establish the fulfilment of {\textbf{RQ3 so we can perceive change in sentiment related to the financial event through time.}}

\begin{table}[]
\centering
\caption{\label{Table7} Example of tweets in `tesco' and \$TSCO' dataset and their corresponding polarities.}
{\small
\begin{tabular}{lll}
\hline
\rowcolor[HTML]{C0C0C0} 
\multicolumn{1}{|l|}{\cellcolor[HTML]{C0C0C0}\textbf{Polarity}} & \multicolumn{1}{c|}{\cellcolor[HTML]{C0C0C0}\textbf{‘tesco’ dataset}}                                                                & \multicolumn{1}{c|}{\cellcolor[HTML]{C0C0C0}\textbf{\$TSCO}}                                                                                 \\ \hline
Positive                                                        & 
\begin{minipage}{0.4\linewidth}
	
“Tesco merging with Londis ?? f****n love Londis”   
\end{minipage}
 & 
  
\begin{minipage}{0.4\linewidth}     
	                                                                         
“Good chat with Steve Fox,Booker/Tesco merger is great news for independents, better access to banking/payment services/…” \end{minipage}
                  \\ \hline

Negative                                                        & 
\begin{minipage}{0.4\linewidth}
	
“Tesco PLC acquires Booker Group for £3.7B creating the "UK's leading food business" uniting the power of the UK's la…”        
\end{minipage}
        & 
\begin{minipage}{0.4\linewidth}
	
“The Booker merger means that Londis and Budgens now under the Tesco umbrella.”                                                          
\end{minipage}    \\ \hline

Neutral                                                         & 
\begin{minipage}{0.4\linewidth}
	
“Tesco buying Booker Group - disastrous for convenience stores who buy stock from Booker Cash\&Carry at prices competitive to supe…”
\end{minipage}
 & 
\begin{minipage}{0.4\linewidth}
	
“Tesco \& Booker merger (aka takeover): potentially terrible news for independent \#retailers \& small grocery chains. Stakeholders beware!” 
\end{minipage}
\\ \hline
\end{tabular}}
\end{table}

\section{Geographical impact of the action}
\label{sec:9}

Although Twitter is one of the most used data sources in data mining, the geo-location component of Twitter is not comparable to other data sources which we can refer to as Location-based social networks. In fact, according to \cite{Morstatter2013}, the geo-located tweets returned by the Streaming API cover up to 90\% of the geo-located tweets extracted from Firehose API. However, this study also reveals that the number of geo-located tweets is low, being only a 1.45\% of the tweets obtained from Firehose API and 3.17\% of the tweets obtained from Streaming API. The total percentage of Tweets geo-located in the `tesco' dataset is consistent with this previous study (Morstatter, Pfeffer, Liu, \& Carley, 2013),{~ }with a percentage of 4.3\% for all the periods in the experiment. Although the number of tweets in the \$TSCO dataset may be not representative enough, we should remark that the percentage of geo-located tweets in the \$TSCO dataset is 0\%, 1 tweet out of a total of 199. Also, there is not variability of those percentages throughout the periods considered (pre- post- and regular). Beyond the percentage of geo-located tweets that the Twitter APIs return, the variation of the geographical distribution of the tweets due to the financial event deserves to be analysed. \ref{Figure7} shows this distribution and illustrates that there is not much variation if we compare post-announcement with the regular period the same days of the week ({10\textsuperscript{th} Feb 07:00 - 12\textsuperscript{th} Feb 23:59 as defined in Table \ref{Table1}.

\begin{figure}
	\begin{tabular}{cc}
	    \textbf{Post-announcement} &	\textbf{Regular 2-weeks-after} \\
		 27\textsuperscript{th} Jan 07:00 - 29\textsuperscript{th} Jan 23:59 & 10th Feb 07:00 - 12th Feb 23:59 \\
\includegraphics[width=0.5\textwidth]{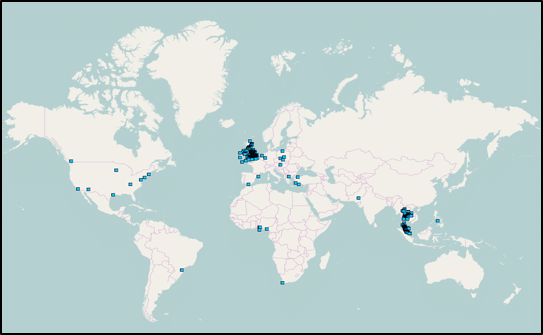}&
\includegraphics[width=0.5\textwidth]{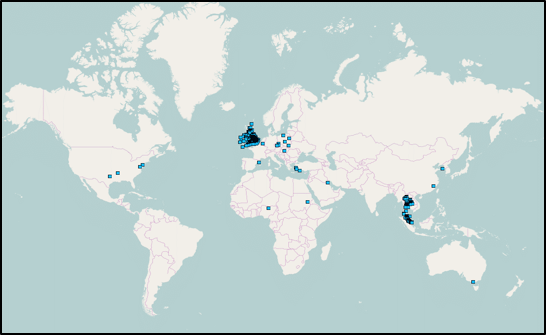}\\
\end{tabular}
\caption{\label{Figure7} Geographical distribution of tweets in `tesco' dataset after the announcement and during a regular period.}
\end{figure}

A deeper inspection of the tweets per country in Table \ref{Table8} confirms that most of tweets come from the countries where Tesco PLC deploy its main business either under Tesco trademark or thorough subsidiary local companies. Apart from UK and Republic of Ireland, the main retail locations of Tesco PLC all over the world are Czech Republic, Hungary, Poland, Slovakia, Turkey, Malaysia and Thailand. According to the results in Table \ref{Table8}, before the announcement, the bigger contribution to Twitter volume corresponds to the UK market which is consistent with the historical roots of the company where its retailing business is fully integrated in the society. Nevertheless, after the announcement, this percentage decreases in favour of other locations over the world, which is a sign of the global impact of the merge so that twitter users outside UK are not so linked to Tesco PLC marketing campaigns during regular period but they are reactive to a relevant event related with a company with presence in their countries. Nigeria is highlighted in Table \ref{Table8} as a country with a definitely high position according to the number of tweets during the post-announcement despite the fact that Tesco does not having presence in this country. 42 of the 43 tweets in Nigeria has the same content but they are tweeted from 42 different users, not being retweets, so that it may be a violation of the spam terms in Twitter rules. With all the above information, we can establish the fulfilment of the {\textbf{RQ4,} so that \textbf{the event impact on Twitter depends on the location and, moreover, the event distorts the geographical distribution of tweets.}

\begin{table}
\caption{\label{Table8} Geographical distribution of tweets in `tesco' dataset}
\includegraphics[width=\textwidth]{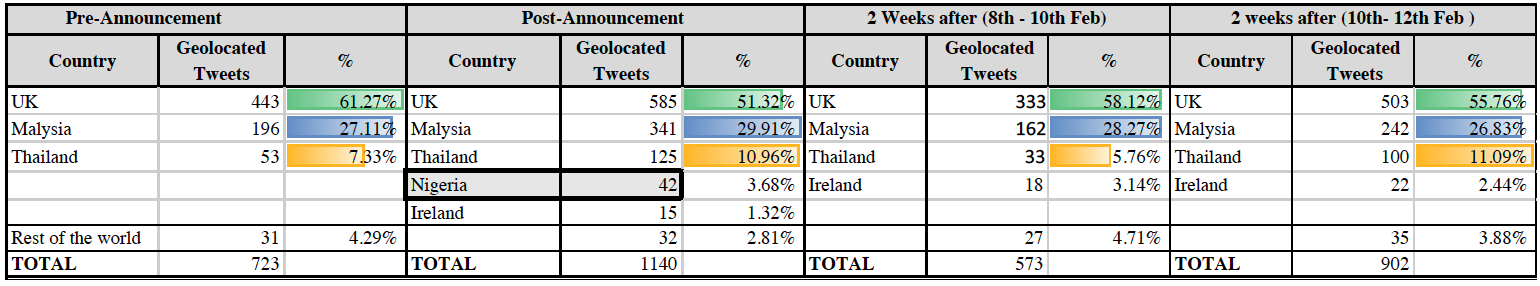}
\end{table}

\section{Rapidness \& Synchronization}
\label{sec:10}

Although our analysis focuses on the permeability of Twitter to financial events, our long-term objective is the use of Twitter as a sensor of irregularities in the stock market. So, this section, includes our findings in the experiment related to the rapidness and synchronisation of Twitter as a channel to the stock market: rapidness in its response to the RNSs of LSE (London Stock Exchange) and synchronisation with the share prices also in LSE.{~ }Regarding the rapidness, the experiment definitively shows the good characteristics of Twitter. The first tweet referring to the RNS was at 7:03 a.m. on 27\textsuperscript{th}, just 3 minutes before the RNS announcement about the Tesco and Booker merge (Figure \ref{Figure8}). Beyond the very first tweet, it is remarkable the rapidness of the peak response to the announcement in both datasets, so that the 27th Twitter time series (`tesco' and \$TSCO) can be considered abnormal time series when a regular Friday is taken as a reference. We highlight that the peak starts from 7:00 to 8:00 both in the \#TSCO and \$TSCO dataset (see Figure \ref{Figure9}).

\begin{figure}
	\begin{tabular}{cc}
\includegraphics[width=0.5\textwidth]{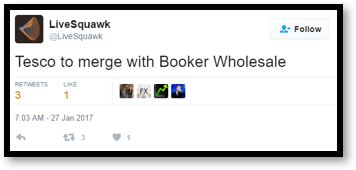} & \includegraphics[width=0.4\textwidth]{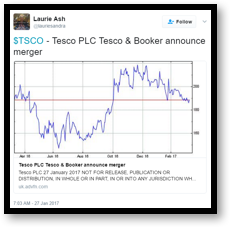}\\
{\small `tesco' dataset: Fri Jan 27 07:03:59 2017} & {\small \$TSCO dataset Fri Jan 27 07:03:46 2017}\\
\end{tabular}
\caption{\label{Figure8} First Tweet referring to the action in the `tesco' and \$TSCO dataset}
\end{figure}

\begin{figure}

	\begin{tabular}{cc}
\includegraphics[width=0.5\textwidth]{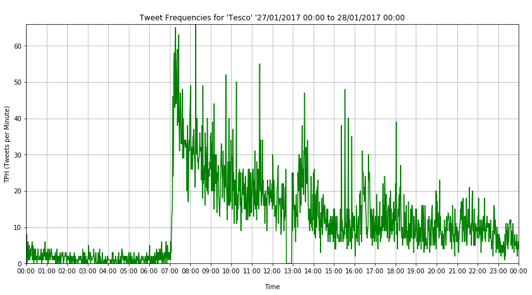} &\includegraphics[width=0.5\textwidth]{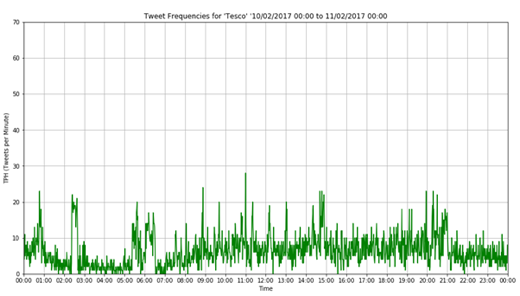} \\
27\textsuperscript{th} January `tesco' &  Regular Friday for `tesco' \\
\includegraphics[width=0.5\textwidth]{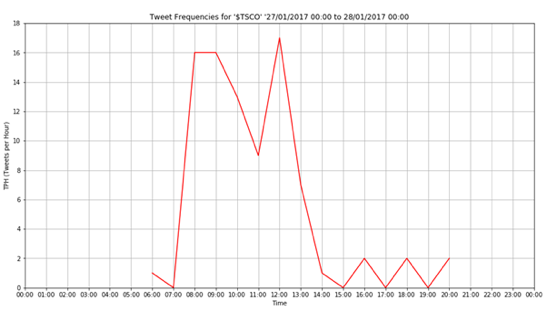} & \includegraphics[width=0.5\textwidth]{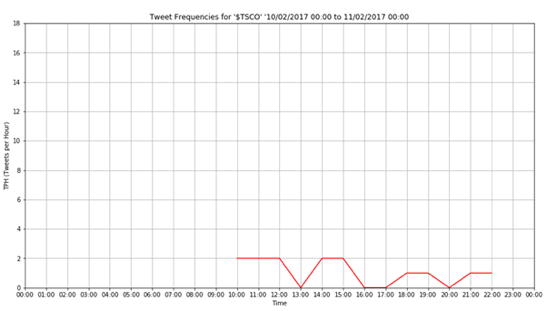} \\
27\textsuperscript{th} January \$TSCO &  Regular Friday for \$TSCO \\
\end{tabular}
	\caption{\label{Figure9} Time series at a minute scale on the 27\textsuperscript{th} January in comparison with a regular Friday.}
\end{figure}

Regarding the synchronisation with the share prices at LSE (Figure \ref{Figure10} and Table \ref{Table9}), it is fair to mention that although the share prices were abnormally low the day before the announcement, we haven't found any reference to the financial vocabulary considered for the action during this period. Moreover, we cannot extract a sound finding given to the concatenation of two events with financial impact in Tesco PLC: the legal actions against Tesco PLC overstatement (see Section VII) and the merge of Tesco and Booker.  In terms of rapidness and synchronisation, we can only partially establish the fulfilment of \textbf{RQ5 so that the permeable layer exhibits rapidness in response to the financial market but we cannot conclude a clear synchronisation with the stock share prices.}

\begin{figure}
	\begin{tabular}{cc}
\includegraphics[width=0.5\textwidth]{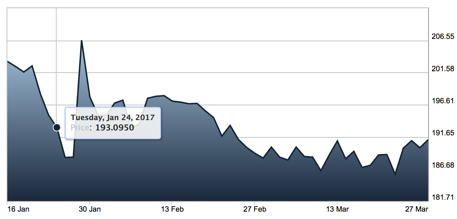} &\includegraphics[width=0.5\textwidth]{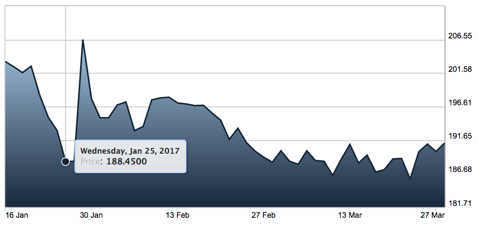} \\
\includegraphics[width=0.5\textwidth]{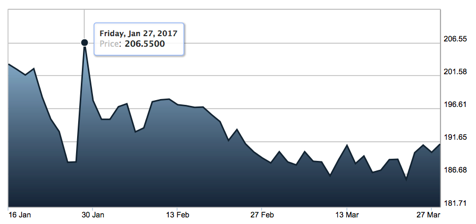} &\includegraphics[width=0.5\textwidth]{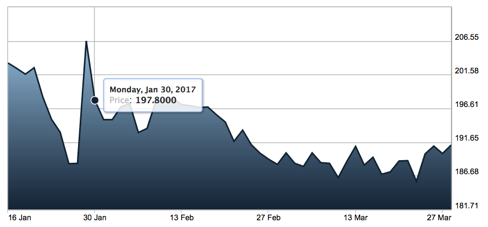}

\end{tabular}	
\caption{\label{Figure10} Main observational points in the Evolution of the Tesco PLC share price (16\textsuperscript{th} January to 27\textsuperscript{th} March 2017).}
	\end{figure}
\begin{table}[]
	\centering
	\caption{\label{Table9} Values of share prices and Twitter volume}
	\begin{tabular}{lllll}
	\hline
	\rowcolor[HTML]{C0C0C0} 
	\multicolumn{1}{|l|}{\cellcolor[HTML]{C0C0C0}\textbf{Period}} & \multicolumn{1}{l|}{\cellcolor[HTML]{C0C0C0}\textbf{Price closing time}} & \multicolumn{1}{l|}{\cellcolor[HTML]{C0C0C0}\textbf{‘tesco’}} & \multicolumn{1}{l|}{\cellcolor[HTML]{C0C0C0}\textbf{\$TSCO}} & \multicolumn{1}{l|}{\cellcolor[HTML]{C0C0C0}\textbf{Total}} \\ \hline
	24th 16:30 to 25th 16:30                                      & 188.45                                                                      & 7,697                                                         & 6                                                            & 7,703                                                       \\ \hline
	25th 16:30 to 26th 16:30                                      & 188.5                                                                       & 8,356                                                         & 4                                                            & 8,360                                                       \\ \hline
	26th 16:30 to 27th 16:30                                      & 206.55                                                                      & 15,295                                                        & 38                                                           & 15,333                                                      \\ \hline
	28th 16:30 to 29th 16:30                                      & ---                                                                         & 10,077                                                        & 8                                                            & 10,085                                                      \\ \hline
	29th 16:30 to 30th 16:30                                      & ---                                                                         & 4,539                                                         & 3                                                            & 4,542                                                       \\ \hline
	\end{tabular}
	\end{table}

\section{Discussion}
\label{sec:11}

This paper inspects the permeability of Twitter to financial events in order to provide evidence which allows Twitter to be used as a social sensor for the financial and stock market. To do that, this permeability should be checked and measured. Bearing in mind that this a single experiment for a single financial event and also that the event had been fully covered by traditional social media, we can conclude that the event in the financial market invaded the Twittersphere on the 27\textsuperscript{th} January 2017, just after the RNS announcement at 7:00, and that the behaviour of the triplet (\$TSCO, \#{[}tesco{]}, ``tesco'') was altered in comparison with the regular behaviour around the company involved in the financial event. At the same time, the experiment shows that other financial events that affected the company (overstatement) during the very same period was only permeable to the more financial oriented tweets and users. As many other social networks, homophily \cite{Choudhury2011} exists in the context of Twitter, that is, the tendency to interact with similar individuals where this similarity may be in several dimensions such as age, location, and occupation, etc. The experiment in this paper shows that the high impact of the merge action cross the fragile boundary between users with specific financial interests (highly connected to each other) to the general audience in Twitter. However, this work has not studied the profile of the users in the general audience which are captured by the Tesco merge, or, to put in another way, to which kind of homophilic words the information flows out from the very experts in the stock market.

Coming back to the fundamental research question addressed in this paper, we sustain and successfully confirmed that {\emph{``Twitter (although not a specific financial forum) is permeable to financial events and this permeability can be analysed by monitoring (1) the name of companies as a keyword, (2) the Cashtag of the company and (3) the hashtags related to that company.''}The results of the experiment on the}{ announcement about the merge of Tesco PLC and Booker Group PLC on the 27\textsuperscript{th} January 2017 show that the \emph{Twittesphere} is permeable to the financial market dynamics thanks to the tweets of a variety of different contributors.{~ }The merge impacted on all the Tesco-related content in Twitter in terms of volume, having a higher impact on }{\$cashtag-content but even altering the tweets' topics in comparison to regular behaviour (altering \#hashtag dynamics and tweets' content).{~ }Also, with the precautions of a single experiment, we also observed changes in polarity and in the geographical distribution of the contributes through time. Finally, the good freshness characteristics of Twitter as news media is confirmed by the rapidness of response to the RNS announcement.

Therefore, the experiment was successful in confirming that a far-impacting financial event causes disturbance in all the features considered in relation with Twitter permeability: information volume, content and sentiment as well as geographical provenance. Nevertheless, the experiment had a little success in identifying some rumour or sign of the announcement prior to the event. Even considering that the experiment was not deployed over the whole Firehose Twitter data, uncovering rumours before the announcement turns definitively into a hard task, if the spreading of rumours in real life is not mimicked inside Twitter, that means, if the rumour is not in Twitter at all. At this respect, and according to \cite{Liu2016}, social media data can only be generalised to human behaviour when social media provides a representative description of human activity. Twitter is a social media which, at least, exhibits some demographic bias. Moreover, Twitter may be providing a skewed representation of content. Although well-known rumour detection algorithms \cite{Vosoughi2015} \cite{Tafti2016} can be applied to Twitter, an alternative approach can be the fusion of financial information from different data sources in a way that we can mitigate the inevitable bias in a single source, and, at the same time, combine their weaknesses and strengthens in a proper representation of the real financial activity.


\bibliographystyle{elsarticle-num} 

\end{document}